\documentclass[aps,prb,twocolumn,superscriptaddress,floatfix,longbibliography]{revtex4-2}
\usepackage{amsmath,amssymb} % math symbols
\usepackage{bm} % bold math font
\usepackage{graphicx} % for figures
\usepackage{comment} % allows block comments
\usepackage{enumitem}
\setlist{noitemsep,leftmargin=*,topsep=0pt,parsep=0pt}
\usepackage{xspace}
\usepackage{xcolor} % \textcolor{red}{text} will be red for notes
\definecolor{lightgray}{gray}{0.6}
\definecolor{medgray}{gray}{0.4}
\usepackage{hyperref}
\usepackage{siunitx}
\usepackage{tabularx}
\usepackage{graphicx}
\newcommand\ket[1]{\vert{#1}\rangle}
\newcommand\bra[1]{\langle{#1}\vert}
\hypersetup{
colorlinks=true,
urlcolor= blue,
citecolor=blue,
linkcolor= blue,
}
\begin{document}

\title{Topological Phases in Coupled Polyyne Chains}
\author{Nisa Ara}
\email{p20210035@goa.bits-pilani.ac.in}
\author{Rudranil Basu}
\email{rudranilb@goa.bits-pilani.ac.in}
\affiliation{Department of Physics, Birla Institute of Technology and Science Pilani, Zuarinagar, Goa 403726, India}

\begin{abstract}

%Coupled Polyyne chains realised in two distinct stacking show new paradigm for one dimensional topological insulators. 
We study the electronic properties of coupled parallel Polyyne chains in a couple of symmetric stacking arrangements, namely the AA stacking and the stacking AB, with the single and triple carbon bonds of one chain aligned (AA) and anti-aligned (AB) with those of the other chain.% with the effect of longitudinal sliding. 
Both these arrangements described by tight binding Hamiltonians, whose parameters are calibrated by matching low energy dispersion provided by first principle calculations, fall in the BDI class of topological classification scheme.
We calculate the topological invariants for all the three topological phases of the system: one for the AA stacking and 2 for the AB one. In the AA stacking, both the insulating and the metallic phase belongs to the same topological phase. Whereas, the model exhibits two different values of the topological invariant in the two different insulating phases (structurally differentiated by transverse strain). In this later stacking though transition between two distinct topological phases with the closure of the gap is practically unachievable due requirement of high amount of transverse strain. We also show existence of 4 non-zero energy edge modes in the AA stacking and that of 2 zero energy edge modes in one of the topological phases for the AB stacking.
\end{abstract}

\maketitle

\section*{Introduction}
Being one dimensional, the Su-Schrieffer-Heeger (SSH) model probably is the simplest system with crystalline symmetry, showing non-trivial, gapped topological phases \cite{su1979solitons,su1980soliton, heeger1988solitons}. Real systems, ranging from the original hydro-carbon polyacetylene chain, the pure carbon Polyyne to the artificially prepared array of photonic crystals or array of cold atoms can be modelled by the SSH Hamiltonian \cite{kitagawa2012observation,atala2013direct,nakajima2016topological,lohse2016thouless,xie2018second}. For polyacetylene, the two different topological phases are realized via two different dimerizations\cite{asboth2016short,li2014topological,delplace2011zak}. Controlled by two alternate hopping parameters, the phases are distinguished by their ratio, whereas the critical point corresponds to the ratio to be unit and closure of the band gap in its electronic structure \cite{hasan2010colloquium,qi2011topological}.

Recently, richer structures in bulk conductivity of a system of a coupled Polyyne chains were predicted via first principle calculations, supported by tight binding computation \cite{basu2022structural}. For realistic spatial separation between the chains, the system becomes conducting for a highly symmetric stacking (we call that AA, keeping in mind the convention used in multi-layer graphene literature\cite{charlier1992first}), caused by inter-chain van der Waals coupling. However, a drastic gap opening happens as one slides one chain with respect to the other, hence breaking particle hole symmetry. Eventually the particle hole symmetry is restored if the sliding is large enough and one gets another commensurate stacking (AB). This particular stacking, in the context of polyacetylene, can be thought of as a coupled chain with altered dimerization. For AB stacked Polyyne ladder, we perform a first principle calculation matching against a tight binding one for low energy states of bands near the Fermi level.

Topologically non-trivial Hamiltonians are classified in 10 classes based on the discrete symmetries - time reversal ($T$), charge conjugation ($C$) and chiral symmetry ($S$). SSH chain %possesses
are symmetric under all these transformations and falls in the BDI class with $T^2 = C^2 = 1$ and $S = TC$\cite{kitaev2009periodic,chiu2013classification}. The coupled Polyyne chain system in AA and AB stackings are described by drastically different coupled SSH Hamiltonians, despite both enjoying the $T,C, S $ symmetries and falling in the 1-dimensional BDI class. The various topological phases within the 1-dimensional BDI class are characterized by the first fundamental group of the space of Hamiltonians and these are integer valued, succinctly captured by the winding number known as Zak phase \cite{zak1989berry}. For chains with open boundary condition, presence of modes of the Hamiltonian localized at the edge of the system is another way to probe topologically non-trivial nature of a system \cite{ryu2002topological,lang2012edge}.

In the present work we characterize the various topological phases of the coupled Polyyne chains for the AA and the AB stacking. We observe that irrespective of the distance between the chains in the AA stacking, and hence oblivious to the fact whether the system is in metallic or insulating, the Zak phase remains fixed. We also notice finite energy edge modes in the insulating regime of AA stacking case. However, in the AB stacking, the system remains insulating with a gap-closure taking place at certain proximity of the chains and again getting back to the insulating phase for further decrease in the inter-chain distance. These two insulating phases have two different Zak phases. Zero energy edge modes are present in that topological phase in AB stacking, which is characterized by larger separation and hence weaker inter-chain van-der Waals coupling. However, the zero-energy modes disappear in the other phase, which can be accessible by strong inter-chain coupling and hence by bringing the chains too close (in comparison to the Polyyne bond lengths).

The organization of the paper is as follows. In section \ref{model}, we discuss about the model, demonstrating with the coupled chain system in both the AA and AB stacking, write their corresponding Hamiltonian and energy dispersion relations. We show the matching of the tight binding results with ab-initio calculation for the AB case here, at the level of dispersion relation. In section \ref{zak}, we present the Zak phase calculations. In section \ref{edge}, we present analytical calculations of the edge modes.% and that in the AA stacking the edge modes.% are not really zero modes, unlike the AB case.

%In this work we calculate the topological index for the metallic and semiconducting phase.\\Topological properties of the solids are characterized by berry phase acquired by the Bloch bands. For 1D system this phase is named as Zak Phase. If there is a quantum mechanical phase transition, the change in Zak phase is uniquely defined which subjects to two topologically distinct phas                   es.

%The electronic properties of coupled Polyyne chains have been studied before \cite{intro_r4}. They exhibit metallic behaviour in AA stacking. On sliding on of the chain in either direction results in metal to semiconductor transition. 
%As shown in original work, perfectly AA stacked carbon coupled chains could be classified in BDI topological class as the Hamiltonian  has time reversal, charge conjugate and chiral symmetry. \\
%This motivates us  to study the topological properties of the system that is calculation of topological invariants and its underlying edges states. We also study the effect of sliding. 

\section{The model}\label{model}
A single carbon Polyyne chain is composed of carbon atoms with alternate triple and single bonds \cite{rusznyak2005bond,lambropoulos2017electronic,al2014electronic,kartoon2018driving}. Single Polyyne chain has a unit cell, consisting of two sublattices $a (\tilde{a})$ and $b(\tilde{b}).$ The lattice spacing is $A$. We consider unequal nearest neighbour (NN) hopping amplitudes with equal site separation length which is an off-set to bond length alternation in the chain. To work with two coupled carbon chains we also introduce inter-chain hopping amplitudes (NN only) which are effectively due to van der Waals interactions. The two symmetric stackings of coupled carbon chains are shown in the Fig.~\ref{chain}(a) and (b). The unit cell has four lattice sites, and lattice period as $A$. In the AA stacking, the triple bonds and the single bonds of the chains are aligned next to each other, whereas the pattern is altered in the AB stacking. These models are comparable to coupled SSH chains or the SSH ladder in \cite{padavic2018topological,kurzyna2020edge,jangjan2020floquet}. 
%\newpage
%\pagebreak
\subsection{AA Stacking}
\begin{figure*}[ht]
    \centering \includegraphics[width=1.8\columnwidth]{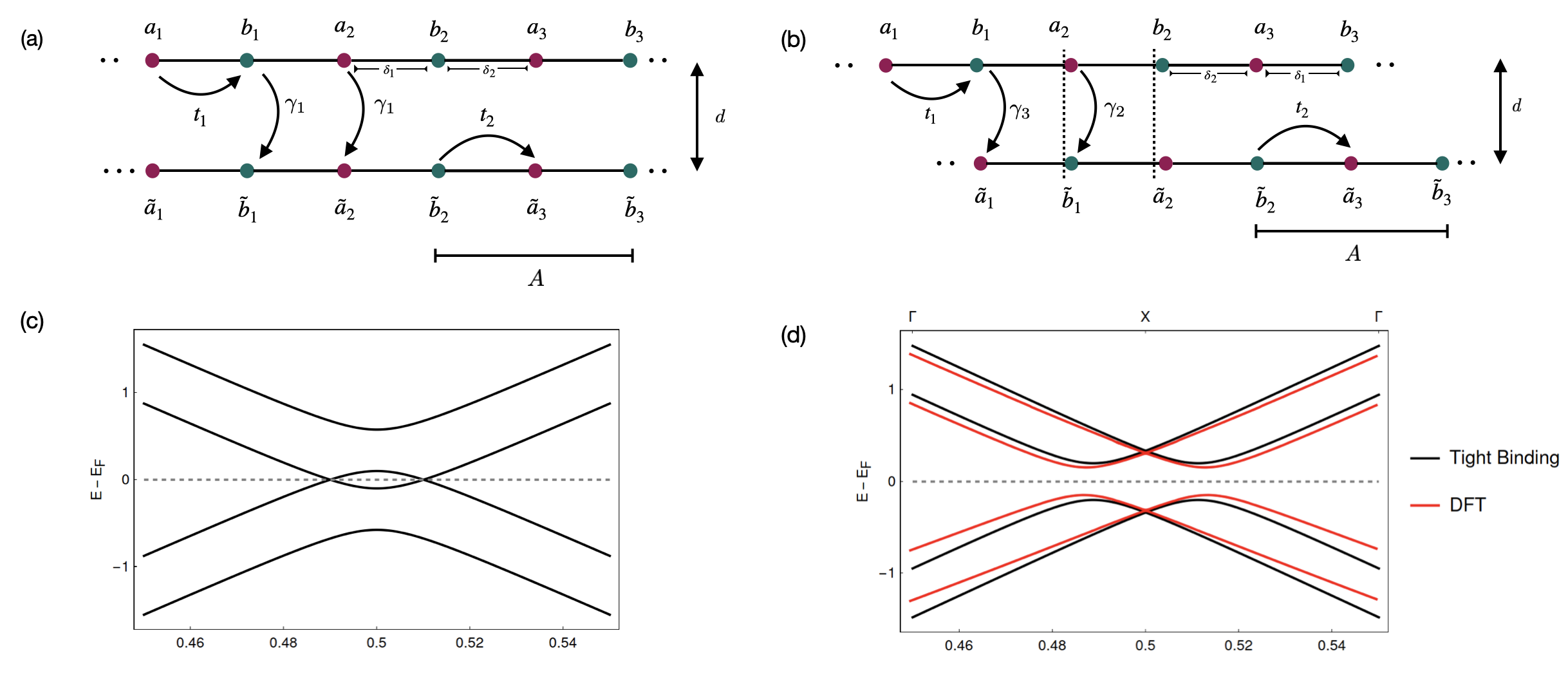}
    \caption{\small Carbon coupled chains (with alternate bond lengths $\delta_1=1.298 \mathring{\mathrm{A}}$ and $\delta_2=1.262 \mathring{\mathrm{A}}$).  in (a) AA stacking, 
	(b) for AB stacking we slide the lower chain (towards left for) for slid distance $\delta_1$. Bulk band structures for (c) AA stacking, (d) AB stacking.}

    \label{chain}
\end{figure*}

In Fig.~\ref{chain}(a)  we have two carbon chains in perfect AA stacking. The intra-chain hopping parameters are $t_1=3.682$ eV and $t_2=3.92$ eV as obtained from the first principle calculations \cite{al2014electronic,basu2022structural}. The tight binding Hamiltonian for this system is  
\begin{align}
    \mathcal{H}_{AA}=&t_1 \sum_{p}  a^{\dagger}_p {b}_p+t_2\sum_{p} a^{\dagger}_{p+1} b_p + t_1 \sum_{p} \tilde{a}^{\dag}_p \tilde{b}_p+\nonumber \\
    &t_2 \sum_{p} \tilde{a}^{\dag}_{p+1}  \tilde{b}_p+ \gamma_1 \sum_{p} (\tilde{a}^{\dag}_{p} a_p+\tilde{b}^{\dag}_{p} b_p)+ h.c.
    \label{AA_h}
\end{align}
Apart from time reversal and charge conjugation, Hamiltonian has on-site chiral symmetry, i.e. under transformations $a \rightarrow -a$ and $\tilde{b} \rightarrow -\tilde{b}$, $\mathcal{H} \rightarrow - \mathcal{H}$.
The tight binding Hamiltonian in Fourier space is
\begin{equation}
 \mathcal{H}_{AA}=  \sum_k {\Psi^{\dag}}_k
 \begin{pmatrix}
    0 & f &\gamma_1 & 0 \\
    f^{*} & 0 & 0 & \gamma_1 \\
    \gamma_1 & 0 & 0 & f \\
    0 & \gamma_1 & f^{*} & 0 \\
    \end{pmatrix}
    \Psi_k,
    \label{AA_hk}
    \end{equation}
where $\Psi^{\dag}_k=( c^{\dag}_k \hspace{0.4cm} d^{\dag}_k \hspace{0.4cm} \tilde{c}^{\dag}_k \hspace{0.4cm} \tilde{d}^{\dag}_k )$ is a four-component fermion and $f=t_1+t_2 e^{-iAk} $. The four energy bands are given by, $\epsilon= \pm |f| \pm \gamma_1$. 
We write the Hamiltonian kernel in terms of Pauli-matrices 
\begin{eqnarray}
H_{AA} = \gamma_1 \sigma_x \otimes \mathbf{1} + \mathbf{1} \otimes (\mathfrak{Re}(f) \sigma_x + \mathfrak{Im}(f) \sigma_y).
\label{aa_pauli}
\end{eqnarray}

The inter-chain NN hopping parameter has been modelled to acquire a gaussian dependence \cite{kariyado2019flat} on the separation of the chains $d$ ,
\begin{equation}
    \gamma_1(d)=t_0 \exp\left(-\frac{d^2}{ \kappa_0{d_0}^2}\right),
    \label{distance}
\end{equation}
here $d_0 = 3.78 \mathring{\mathrm{A}}$ is the relaxed inter-chain separation and $\kappa_0= 0.4$ is the gaussian width controlling parameter and  $t_0 = 3.063$ eV gives best match for energy bands to that of obtained from first principle calculations \cite{basu2022structural}. The system is insulating for $\gamma_1<|t_1-t_2|$, whereas for $\gamma_1 \geq |t_1-t_2| $ the lowest energy conduction and the valence bands cross and we find couple of nodal points as shown in Fig.~\ref{chain}(c).  
\subsection{AB stacking}\label{ab_model}
We obtain AB stacking by sliding one of the chains by keeping them parallel to obtain opposite dimerization as shown in Fig.~\ref{chain}(b). Since we are considering only inter-chain NN hopping, couplings between different sub-lattice sites contribute. A very similar stacking has been discussed in \cite{luo2022topological,li2017topological} as coupled SSH chains (ladder system).
The tight binding Hamiltonian for this system is 
\begin{align}
    \mathcal{H}_{AB}=   &t_1 \sum_{p} a^{\dagger}_p {b}_p+ t_2\sum_{p} a^{\dagger}_{p+1} b_p+\nonumber \\
    & t_1 \sum_{p} \tilde{a}^{\dag}_p  \tilde{b}_p+t_2\sum_{p}  \tilde{a}^{\dag}_{p+1}  \tilde{b}_p+\nonumber \\
    &\gamma_2 \sum_{p} a^{\dag}_{p+1} \tilde{b}_p+\gamma_3\sum_{p}   \tilde{a}^{\dag}_{p} b_p+ h.c.
    \label{AB_h}
\end{align}
AB stacking preserves all the symmetries that are in AA case (Hamiltonian has on-site chiral symmetry under transformations $a \rightarrow -a$ and $\tilde{a} \rightarrow -\tilde{a}$). The tight binding Hamiltonian in Fourier space is (considering $\gamma_2 = \gamma_3$). 
\begin{equation}
\mathcal{H}_{AB}=  \sum_k {\Psi^{\dag}}_k
 \begin{pmatrix}
    0 & f & 0 & \gamma_2g \\
    f^{*} & 0 & \gamma_2 & 0 \\
    0 & \gamma_2 & 0 & f \\
    \gamma_2{g}^{*} & 0 & f^{*} & 0 \\
 \end{pmatrix}
\Psi_k.
\label{AB_hk}
\end{equation}
where $g=e^{-iAk}$. In terms of Pauli-matrices the Hamiltonian kernel in \eqref{AB_hk} is,
\begin{align}
    H_{AB}=& \gamma_2( (1+\mathfrak{Re}(g) )\sigma_x \otimes \sigma_x + \mathfrak{Im}(g) \sigma_x \otimes \sigma_y+ \nonumber \\
    & \mathfrak{Im}(g) \sigma_y \otimes \sigma_x+(1-\mathfrak{Re}(g)) \sigma_y \otimes \sigma_y) +\nonumber \\
    &\mathfrak{Re}(f) \mathbf{1} \otimes \sigma_x + \mathfrak{Im}(f)\mathbf{1} \otimes \sigma_y.
    \label{AB_hk1}
\end{align}
\begin{figure}
    \centering
    \includegraphics[width=0.48\textwidth]{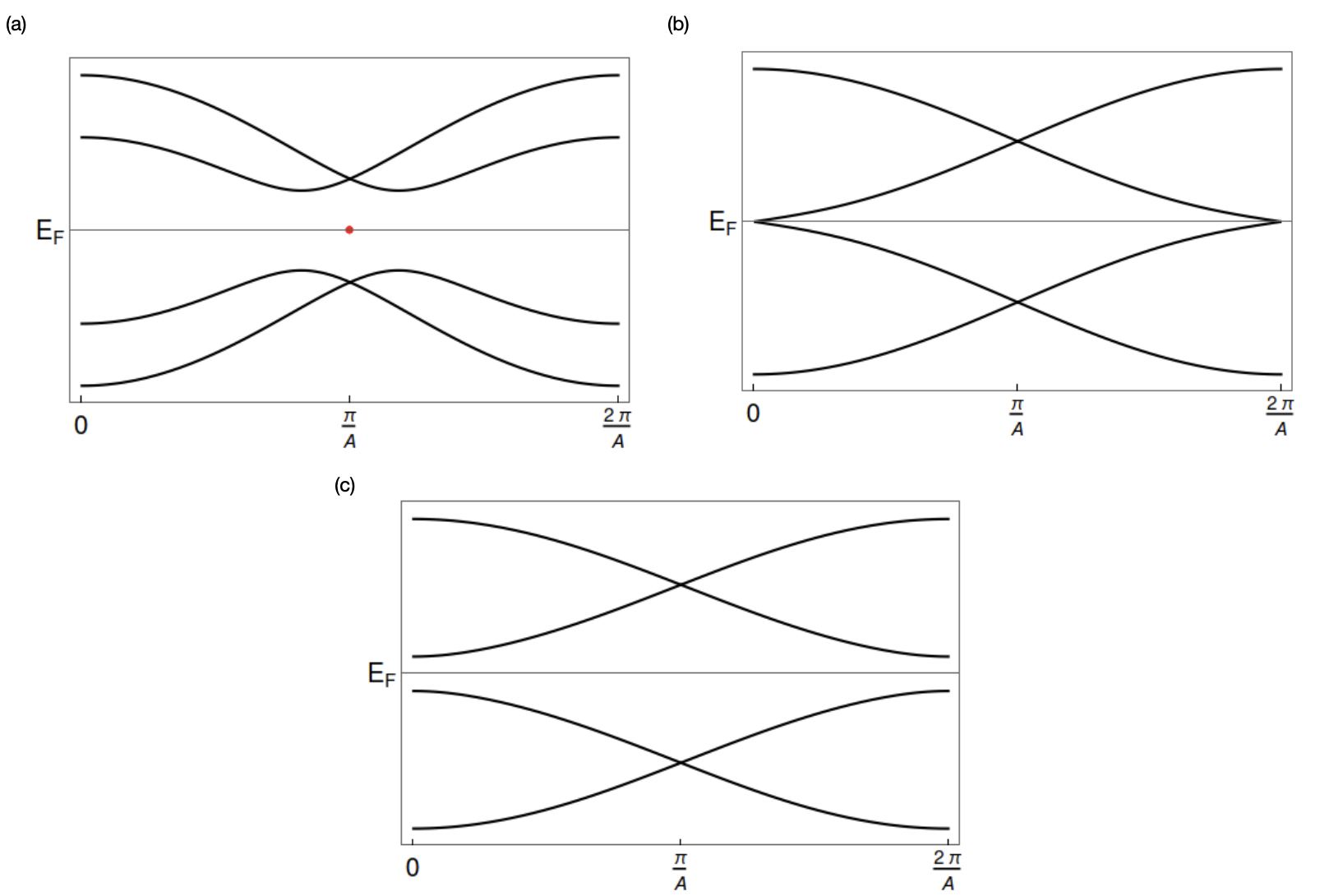}
    \caption{\small Energy bands for the AB stacking (a) $\gamma_2<|t_t+t_2|$. As explained in the section \ref{ab_model}, there are two edge modes. The cases for (b) $\gamma_2= |t_1+t_2|$ and  (c) $\gamma_2>|t_1+t_2|$ are not realizable for the coupled Polyyne chain because that will require bringing the chain too close, as explained below. The closing and opening up of the band gap clearly signals a topological phase transition in this model.}
    \label{AB_bands2}
\end{figure}
We get four bands $\epsilon'$=$\pm \sqrt{{|f|}^2+{\gamma_2}^2\pm 2\gamma_2(t_1+t_2)\cos{|\frac{Ak}{2}}|}$. In AB stacking as shown in Fig.~\ref{chain}(b)  $\tilde{a}_p$ is at very small slide distance ($\delta_1-\delta_2$) with respect to $b_p$ and $a_{p+1}$ coincides with $\tilde{b}_p$. From the point of view of first principle calculations this is a small alteration in the position of one of the  sub-lattice sites (in one  of the chains). We obtain the band structure for inter-chain distance $d=3.55 \mathring{\mathrm{A}}$ to compare with the same obtained for AA stacking and system is insulating as shown in Fig.~\ref{chain}(d). 

According to the energy spectrum obtained from \eqref{AB_hk},  the system is in an insulating phase for  $\gamma_2 < (t_1+t_2)$ and $\gamma_2>(t_1+t_2) $ where transition happens for $\gamma_2=(t_1+t_2)$. The band gap closes at the center of the first Brillioun zone exhibiting semi-metallic behaviour.
As in the AA stacking case, we assume the inter-chain hopping parameter $\gamma_2$ to have the same gaussian dependence as described in \eqref{distance}. However for large $\gamma_2$, such that $\gamma_2 > t_0$ (the values are comparable to that calculated for the previous model), inter-chain distance obtained in such case is not real; since $d^2= -\kappa {d_0}^{2} \ln{|\frac{\gamma_2}{t_0}|}$. Therefore semi-metallic and insulating phases observed for $\gamma \geq (t_1+t_2)$ are not physically achievable with the same Gaussian model.
%We study the effect of varying inter-chain distance which is equivalent to transverse strain to the system. 
%We observe insulating phase for $\gamma_2<|t_t+t_2|$. For $\gamma_2= |t_1+t_2|$ band gap closes, exhibiting semi-metallic behaviour. The gap reopens for $\gamma_2>|t_1+t_2|$, however such system is not physically achievable.
%\textcolor{red}{Explain why this is not physically achievable}.
\subsubsection*{Computational Details}
\label{cp}
We obtain the tight-binding parameters for AB stacking by density functional theory (DFT) technique based on the PBE implemented in  Quantum espresso package \cite{giannozzi2009quantum}. We fix the inter-chain distance as 3.55 $ \mathring{\mathrm{A}}$.
The Brillouin zone is sampled with k-point grids 12 × 1 × 1 and the basis set is plane waves with a maximum kinetic energy of 250 eV. We plot two energy bands in the proximity of the Fermi level and match with that of obtained from tight binding calculations and we get $t_1=3.78$ eV and  $t_2=3.92$ eV. The above mentioned first principle DFT calculations predict the existence of the kinks/ nodal points away from the Fermi level as in the Fig.~\ref{chain} (d). The existence of the kinks necessitates $\gamma_2 = \gamma_3$. Matching for the Fermi velocity and the location of the kinks, from the bands obtained from the tight binding kernel, then gives $\gamma_2=\gamma_3 = 0.27 \, \, \mathrm{eV}$.
%\textcolor{red}{Make the following table in landscape}
\begin{table} 
\begin{center}
\begin{tabular}{ | c | c | c | c | c | c | } 
\hline
  {(eV)} & $t_1$  & $t_2$ & $\gamma_1$ & $\gamma_2$ & $\gamma_3$ \\
%  {} & (eV) & (eV) \\
 \hline
   AA   &   3.682  & 3.92 & 0.37  &0  &  0   \\
   \hline 
   AB   &   3.78  & 3.92 & 0  & 0.27  &  0.27   \\
   \hline
\end{tabular}
\caption{ \small Tight binding parameters, used for the determination of  topological invariants and edge modes.}
\label{table}
\end{center}
\end{table}
\section{Topological invariants} \label{zak}
Here we calculate the Zak phase for the system in two distinct stacking as described in the earlier section. We sum over Zak phases obtained for occupied bands:  \cite{midya2018topological,lee2022winding} 
\begin{equation}
    \Phi= i \int_{  \frac{-\pi}{A}}^{\frac{\pi}{A}} \sum_{i \in \mathrm{occupied}} \bra{X_i} \dfrac{d}{dk}\ket{X_i} dk.
    \label{zak1}
\end{equation}
Here $|X \rangle$ are the $k$ dependent eigenvectors of the Hamiltonian kernel.
\subsection{AA Stacking}
The ortho-normal eigenvectors of the Hamiltonian kernel $H_{AA}$ in  \eqref{AA_hk} describing AA stacking are
\begin{equation}
    \ket{X_{1,2}^{AA}}=
    \frac{1}{2}\begin{pmatrix}
    \pm  e^{-i\theta_k}\\
    \pm 1\\
    e^{-i\theta_k}\\
    1\\
    \end{pmatrix},
    \ket{X_{3,4}^{AA}}=
   \frac{1}{2} \begin{pmatrix}
    \mp  e^{-i\theta_{k}} \\
    \pm 1\\
    - e^{-i\theta_k}\\
    1\\
     
    \end{pmatrix},
    \label{e31}
\end{equation}
where
\begin{equation}
    \theta_k =\tan^{-1}\left(\frac{t_2 \sin(kA)}{t_1 + t_2 \cos(kA)}\right).
\end{equation}
The Zak phase calculated as sum of the contributions from the occupied bands labelled as 3,4 is quantized as 2$\pi$. Interestingly, the Zak phase is not affected by $\gamma_1$. To understand this, we can view the Hamiltonian kernel \eqref{aa_pauli} as the Hamiltonian of a couple of spin 1/2 particles, the first one being subject to a constant magnetic field $\mathbf{B_1} \sim (\gamma_1,0,0)$ and the second one in a varying (here the crystal momentum $k$ plays the role of time) one $\mathbf{B_2} \sim (\mathfrak{Re}(f),\mathfrak{Im}(f),0)$. The second spin acquires a Berry (Zak) phase in adiabatic completion of one cycle $0< k \le 2\pi /A$, while the second one doesn't, it being affected by a constant magnetic field. Hence whatever Zak phase the system acquires through a full traversal of the Brilloiun comes from the contribution of the second spin and is not impacted by $\gamma_1$. Therefore the Zak phase is same as that of a single SSH chain with the particular dimerization of $t_2> t_1$.

As $\gamma_1$ increases, starting from the insulating phase, with the the chains being brought closer, at certain point the gap closes and any further increase of $\gamma_1$ doesn't open up the gap. This is also a reconfirmation of the fact that the point $\gamma_1 = |t_1 -t_2|$ is not a topological phase transition point.
\subsection{AB stacking}
%For any  intermediate sliding, such that $\gamma_2$ and $\gamma_1$ are non zero, and $\gamma_1 <|t_1-t_2|$, 
%the Zak phase is not quantized.
As we slide the chain for a small slid distance ($ < \delta_1$) band gap opens and the particle hole symmetry breaks and this system no longer falls under BDI topological class. However the particle hole symmetry is restored for the AB model described in the previous section. Hence we expect quantized values for the Zak phase.

In order to proceed, we use the eigenvectors of the AB stacking kernel corresponding to the occupied bands, as written in the Appendix. 

%\begin{equation}
%    \Phi_{AB_I}= i \int_{\frac{-\pi}{A}}^{\frac{\pi}{A}} \sum_{i=occ} \bra{{{X_i'}}} \dfrac{d}{dk}\ket{{{X_i'}}} dk= 3\pi .
%    \label{ab_zak}
%\end{equation}

Numerical integration yields the Zak phase coming from the sum of the contributions of the occupied bands as:
  \[ \Phi_{AB} =
  \begin{cases}
         3\pi, &  \gamma_2 <(t_1 + t_2)  \\
     2\pi, &  \gamma_2 > (t_1+t_2).\\
  \end{cases}
\] 
It can be seen that depending on the value of $\gamma_2$, topological invariant can be $2\pi$ or $3\pi$ representing different topological phases. 
%Any immediate sliding could be realised with $\gamma_2 \neq \gamma_3$ which results into the removal of the kink; but the system  still holds all the symmetries and is topological.  

\section{Edge Modes}\label{edge}
Unlike in the previous case, where we used periodic boundary conditions (PBC), we will work with open boundary conditions (OBC) to capture the physics of edge modes of the coupled Polyyne chains.
\subsection{Finite edge modes in AA stacking}
 %We consider a finite chain with open boundary conditions, and  numerically diagonalize the Hamiltonian Kernel in \eqref{AA_h} keeping the values for tight binding parameters as obtained from first principle calculations \cite{basu2022structural} and do not find any zero energy modes.

%\begin{figure}[!h]
%\centering
%\begin{subfigure}{0.48\textwidth}
%    \includegraphics[width=\textwidth]{edge1.png}
%    \caption{}
%\end{subfigure}
%\hfill
%\begin{subfigure}{0.48\textwidth}
%    \includegraphics[width=\textwidth]{edge2.png}
%    \caption{}
%\end{subfigure}
%\caption{Here we have taken 2000 atoms per chain, $t_1=3.628$, $t_2=3.92$ as reported for Polyyne chain (a) $\gamma_1=0$ (b)$\gamma_1=0.05$}
%\end{figure}
\begin{figure} 
\centering  \includegraphics[width=0.48\textwidth]{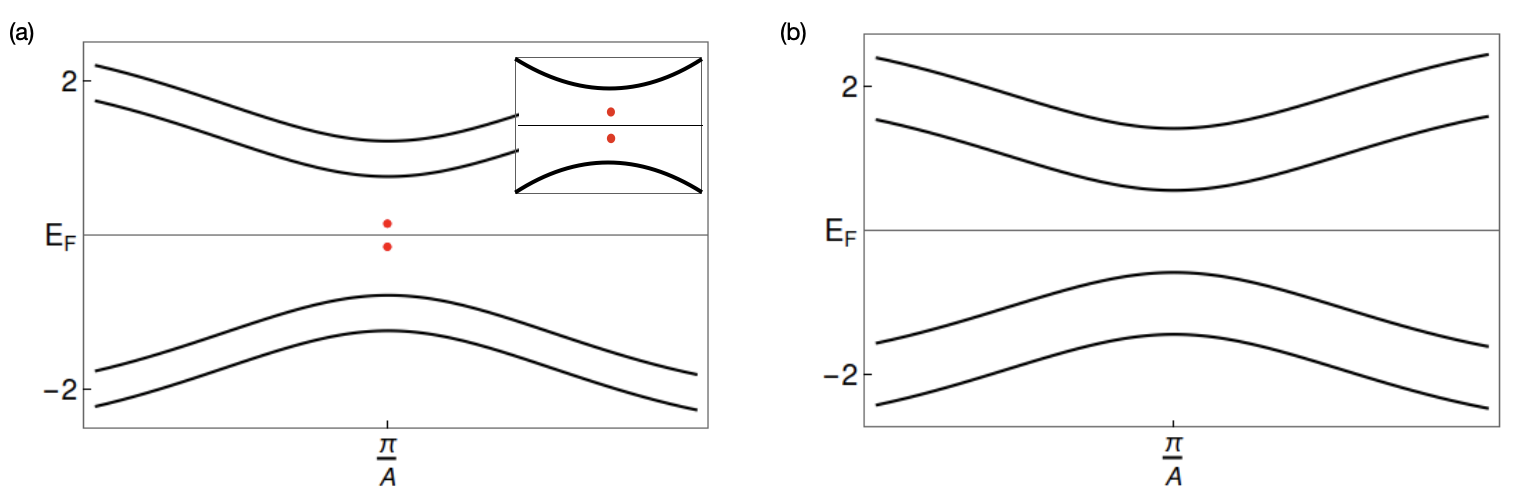}
\caption{\small(a) There are four edge modes with energies $\pm \gamma_1$ isolated from the bulk for $\gamma_1 < \frac{|t_1-t_2|}{2}$, isolated from the bulk of the spectrum; 
(b) edge modes merge with the bulk for $\gamma_1 \geq \frac{|t_1-t_2|}{2}.$}
\label{chain_1}
\end{figure}
To start-off with, we consider a finite chain and using the parameters found by DFT calculations, we found finite energy isolated modes, which we first describe below.

For the AA stacked case, let's first focus on the situation, where $\gamma_1=0$, ie. two decoupled Polyyne chains with OBC. It is well known that a single SSH chain has a couple of zero energy edge modes. Hence it is expected that the decoupled system will have four of them. After turning on finite $\gamma_1$, but still keeping the system insulating, i.e.  $\gamma_1<|t_1-t_2|$, we find modes of energy exactly equal to $\pm |\gamma_1|$, two are $+|\gamma_1|$ and other two are $-|\gamma_1|$ isolated from the bulk. Increase in $\gamma_1$ makes their energies move away from the Fermi level and merge with the bulk spectrum for $\gamma_1 = \frac{|t_1-t_2|}{2}$. This is displayed in the Fig.~\ref{chain_1}.

As we have seen for the PBC, since the Zak phase does not change with the inter chain hopping parameter, we expect that the existence of the edge modes shouldn't be affected by the value of $\gamma_1$.

In order to understand this phenomena analytically and to see that these finite energy modes are indeed localized at the edges, we write the basis which diagonalizes the Hamiltonian kernel in \eqref{AA_hk} as $\psi^{\dag}_k=( \alpha^{\dag}_k \hspace{0.4cm} \beta^{\dag}_k \hspace{0.4cm} \tilde{\alpha}^{\dag}_k \hspace{0.4cm} \tilde{\beta}^{\dag}_k  )$, where $\alpha^{\dag}_k= \frac{1}{2}(e^{-i\theta_k} c^{\dag}_k-d^{\dag}_k -e^{-i\theta_k} \tilde{c}^{\dag}_k+ \tilde{d}^{\dag}_k)$. Consider a finite chain for $P$ unit cells, such that the spatial modes $b_0, \tilde{b}_0$, $a_{P+1}$, and  $\tilde{a}_{P+1}$ are not accessible states to the system\cite{delplace2011zak}. Using $c^{\dagger}_k=\frac{1}{\sqrt{P}}\sum_{p} e^{ikAp}a^{\dagger}_{p}$ etc. we have
\begin{multline}
    \alpha^{\dagger}_k=\frac{1}{2\sqrt{P}} \sum_{p} ( e^{i(kAp-\theta_k)} a^{\dag}_p- e^{(ikAp)}b^{\dag}_p- \\
    e^{i( kAp-\theta_k)} \tilde{a}^{\dag}_p+e^{(ikAp)}
    \tilde{b}^{\dag}_p ).
    \label{e1}
\end{multline}
We can invert the above equation and express the real space oscillators in terms of $\alpha_k$'s. However, in order to achieve the boundary conditions of having no finite excitation arising from the $b_0$ etc. modes, it is better to choose a different basis consisting of only real sinusoidal coefficients rather than the complex phases. This can be done by choosing a linear combination of $\alpha^{\dagger}_k$ and $\alpha^{\dagger}_{-k}$ leading to the new basis of modes as  $\epsilon^{\dagger}_k=\frac{1}{\sqrt{2}}(\alpha^{\dagger}_k-\alpha^{\dagger}_{-k})$. Now \eqref{e1} becomes 
\begin{multline}
    \epsilon^{\dagger}_k=\frac{i} {\sqrt{2P}} \sum_{p} [\sin{(kAp-\theta_k)} a^{\dag}_p- \sin{(kAp)}b^{\dag}_p- \\
    \sin{( kAp-\theta_k)} \tilde{a}^{\dag}_p+\sin{(kAp)}
    \tilde{b}^{\dag}_p ].
    \label{e2}
\end{multline}
If the modes whose energies are isolated from the bulk spectrum have to qualify as edge modes, their corresponding labelling in Fourier space must be given by complex wave-number $k=k_0 +i \lambda$. Finite $\lambda$ results in a new scale $\zeta = 1/\lambda$, which in turn is the localization depth of the isolated modes in the bulk of the chain. This phenomenon is similar to the attenuation of the electromagnetic wave at the surface of the conductor. Substituting $k=k_0 +i \lambda$ in \eqref{e2}, we get: %If we think of localization length of the edges states as skin depth of the conductor then the solution for k takes the form $\tilde{k}=k_0+i\lambda$, where $\lambda=\frac{1}{\zeta}$,
\begin{multline}
    \epsilon^{\dagger}_k=\frac{i} {\sqrt{2P}} \sum_{p}
    (\nu_k a^{\dag}_p
    -\sin{[(k_0+i\lambda)Ap]}             b^{\dag}_p- 
    \nu_k  \tilde{a}^{\dag}_p+\\
    \sin{[(k_0+i\lambda)Ap]}              \tilde{b}^{\dag}_p 
    ),
    \label{e3}
\end{multline}
where
\begin{equation}
    \nu_k=\frac{ \sin{[(k_0+i\lambda)Ap}]\cos{(\theta_{\tilde{k}}})- 
 \sin(\theta_{\tilde{k}})\cos{[(k_0+i\lambda)Ap}]}{|f(\tilde{k})|},
 \end{equation}
and
\begin{equation}
  |f(\tilde{k})| =\pm t_2 \sqrt{ {\left(\frac{t_1}{t_2}\right)}^2+1+2\left( \frac{t_1}{t_2}\right) \cos{(k_0+i\lambda)A}}.
\end{equation}
We demand the coefficients of $b_0, \tilde{b}_0,a_{P+1}$, and  $\tilde{a}_{P+1}$ in \eqref{e3} to vanish. It follows therefore that
\begin{equation}
  \frac{t_1}{t_2}=\frac{  - \sin{[A(k_0+i\lambda)P}]}{    \sin{[A(k_0+i\lambda)(P}+1)]  }.
\label{e4}
\end{equation}
Since $\frac{t_1}{t_2}$ and $|f(k)|$ are real, we have $k_0=\frac{\pi}{A}$, and \eqref{e4} gives $\frac{t_1}{t_2}=\frac{ \sinh{[ \lambda PA}]}  { \sinh{[\lambda(P+1)A] } }.$
%\begin{equation}
%  \frac{t_1}{t_2}=\frac{ \sinh{[ \lambda PA}]}  { \sinh{[\lambda(P+1)A] } }.
%\end{equation}
This gives the energy of the edge localized mode as: $\epsilon = -\gamma_1 - |f(k_0 + i \lambda)|= -\gamma_1 -t_2\dfrac{ \sinh{[ \lambda A}]}  { \sinh{[\lambda(P+1)A] } }$.
%and we get energy as
%\begin{equation}
%    \epsilon= -\gamma_1 -t_2\frac{ \sinh{[ \lambda A}]}  { \sinh{[\lambda(P+1)A] } }.
%\end{equation}
Existence of $\tilde{k}$ hence proves existence of an edge mode. When $\zeta \ll PA $, %we have $\frac{t_1}{t_2} \approx e^{-\lambda A}$, then 
$|f|= 0$ and therefore edge states are exactly -$|\gamma_1|$ as expected by finite numerical analysis earlier. Similarly we find another edge mode with energy -$\gamma_1$ and two more with energy $+\gamma_1$ by choosing appropriate basis in $\beta^{\dag}_k,\tilde{\alpha}^{\dag}_k$ and $ \tilde{\beta}^{\dag}_k$, as we showed for $\alpha^{\dag}_k$ in \eqref{e1}. Therefore there are four finite edge modes as shown in Fig.~\ref{chain_1}. From the dispersion relations coming from \eqref{AA_hk}, we see that the valance band maximum and conduction band minimum are respectively $\mp \gamma_1 \pm |t_1-t_2|$. A curious take home message from this analysis is that although the edge modes cease to appear as isolated ones in the whole spectrum as $\gamma_1 \rightarrow \frac{|t_1-t_2|_{+}}{2}$, they still are edge modes. 

\subsection{Zero energy modes in AB stacking}
Consider the chains with open boundary conditions, with $P$ unit cells in each chain. To obtain the perfect dimerization, we discard sites with no inter-chain hopping, appearing at the edges due to sliding and we keep only even number of sites in each chain we again discard  two sites from  one of the edges. For example in Fig.~\ref{chain} (b), we discard  $a_1, \tilde{b}_3, b_3$ and $\tilde{a}_3$. This results into $2(P-1)$ sites in each chain. 

In this stacking arrangement, one of the chains is in dimerization as ($t_1<t_2$) which favours the existence of zero edge modes while other chain is in opposite dimerization where no such modes exist. Therefore unlike AA stacking, here we get exactly two zero energy modes. For a finite number of sites, we numerically diagonalize the Hamiltonian kernel \eqref{AB_h} to find a couple of eigenvalues isolated  from the bulk spectrum, for the parameters in the table \ref{table}. These approach zero energy with increasing number of sites as shown in Fig.~\ref{AB_bands2}(a).
%We find two zero energy modes isolated from the bulk spectrum, by numerically diagonalizing the Hamiltonian kernel in \eqref{AB_h} (written in the real basis $\Psi^{\dag}=( b^{\dag}_p \hspace{0.4cm} \tilde{a}^{\dag}_p \hspace{0.4cm} \tilde{b}^{\dag}_p \hspace{0.4cm} a^{\dag}_{p+1} \h.space{0.3cm} ...\hspace{0.3cm} \tilde{b}^{\dag}_{P-1} \hspace{0.4cm} a^{\dag}_{P} )$.)

For more generic aspects and the fates of these zero modes as one tunes $\gamma_2$ we carry out the following analysis. For large $P$ ($P \rightarrow \infty$), we consider that there exists a zero energy mode of the Hamiltonian kernel as shown in \eqref{matrix} in the appendix, with the corresponding eigenvector being% then it must belong the energy spectrum of the Hamiltonian kernel, satisfying the eigenvalue equation. We write the corresponding eigen vector as 
$(  \alpha_1 \hspace{0.3cm} \alpha_2 \hspace{0.3cm} \alpha_3 \hspace{0.3cm} \hdots \hspace{0.3cm} \alpha_{4(N-1)})$. The zero-eigenvalue condition gives us the two following recurrence relations   
\begin{equation}
    t_2 \alpha_{2n+1}+\gamma_2 \alpha_{2n+3}+t_1\alpha_{2n+5}=0 \hspace{0.4cm}\text{for  } n=0,1 \hdots
    \label{odd}
\end{equation}
\begin{equation}
    t_1 \alpha_{2n}+\gamma_2 \alpha_{2n+2}+t_2\alpha_{2n+4}=0 \hspace{0.4cm}\text{for  } n=1,2 \hdots
    \label{even}
\end{equation}
For the sake of simplicity, we label the odd components as $\alpha_{2n+1} \equiv B_n$ and the even components as $\alpha_{2n} \equiv D_n$. 
%In order to collect all $\alpha_{2n+1}$ (odd terms), we write $\alpha_{2n+1}$, $\alpha_{2n+3}$, and $\alpha_{2n+5}$ as  $B_{n}$,  $B_{n+1}$ and $B_{n+2}$ respectively. We consider similar solution $D_n$ for \eqref{even} to collect all $\alpha_{2n}$(even terms). 
We obtain solutions for \eqref{odd} and \eqref{even},
\begin{subequations}
\begin{eqnarray}
    B_n=C_1 \left( \frac{-\gamma_2+\kappa}{2t_1}\right)^n+C_2 \left( \frac{-\gamma_2-\kappa}{2t_1}\right)^n,\\
    \label{ab_1}
    D_n=C_3 \left( \frac{-\gamma_2+\kappa}{2t_2}\right)^n+C_4 \left( \frac{-\gamma_2-\kappa}{2t_2}\right)^n,
\end{eqnarray}
\end{subequations}
where $\kappa=\sqrt{{\gamma_2}^2-4t_1t_2}$ and $C_1,C_2,C_3$ and $C_4$ are undetermined coefficients, out of which, two can be fixed i.e. $C_1 = C_3 =1$, with $C_2$ and $C_4$ remaining undetermined. Since the eigenvector should be normalizable for the zero-mode to exist, we must have the sum
%\begin{equation}
    $\sum_{n=1}^{P} (|B_n|^2+|D_n|^2)$
%\end{equation} 
to converge as $P \rightarrow \infty$. To start-off with we focus on the range ${\gamma_2}^2<4t_1t_2$. % keep the parameters as obtained in section \ref{cp} and solve for ${\gamma_2}^2<4t_1t_2$,
As shown in the appendix \ref{series}, the series converges. On the other hand, if we focus on the case ${\gamma_2}^2>4t_1t_2$, convergence is only achieved for $\gamma_2 <t_1 +t_2$. Hence this condition makes the eigenvector normalizable, substantiating the existence of a zero mode for this condition. Since there are two undetermined constants $C_2, C_4$, the vector subspace of zero modes in 2-dimensional and we get exactly a couple of zero modes, as expected. From the perspective of particle hole symmetry also, the existence of the second zero mode is understandable. The existence of zero energy modes isolated from the bulk necessitates that they are edge modes. 

We note that the system is insulating for $\gamma_2 < t_1 +t_2$, with the gap closing for $\gamma_2 = t_1 +t_2$ and again opening for higher values of $\gamma_2$. This being a topological phase transition, is therefore is reconfirmed by vanishing of the couple of zero edge modes across the transition point.
\section*{Conclusions}
We characterize topological phases of the two coupled Polyyne chains. In AA stacking, the symmetries enjoyed by the system dictates it to be in the BDI topological class and the topological invariant is captured by the quantized Zak phase. The Zak phase is robust under transverse strain and is unaffected as the electronic structure passes from insulating to metallic phase. This behaviour is reconfirmed by the analysis of edge modes. We model the AA stacking coupled chain system out of two single SSH chains in the dimerization having a couple of edge modes. Hence the coupled system possesses four edge modes, which curiously are away from the Fermi level. When the chains are brought close to each other, with the inter-chain van der Waals coupling increasing, the energies of edge modes merge with the bulk spectrum; however the edge localization is still retained.

%system is in non-trivial topological phase which is robust under transverse strain. The edge modes show explicit dependence on inter-chain hopping parameter. There are exactly four zero energy modes if chains are decoupled ($\gamma_1=0$). However when they are brought closer, these edge modes are finite and exactly equal to $\pm \gamma_1$.
The other stacking, ie. the AB one, also falls in the BDI class and hence is expected to have quantized Zak phase. Working with the tight binding model of a system of coupled SSH chains in altered dimerization configuration, we uncovered a couple of different topological phases, both of which are insulating, with the gap closing at the transition point in the space of tight binding parameters. To make sure that this tight binding model is a reliable one for describing AB stacked polyyne chains, we did a match of parameters against the electronic structure given by a first principle calculation. However we note that closing of the gap in AB stacking is practically unfeasible for the polyyne system as this would require high amount of transverse strain. This model, being built up from two differently dimerized SSH chains, has only two edge modes of zero energy in one of the topological phase. These modes disappear after as the gap closes and reopens due to increase of inter-chain van der Waals coupling. 

%, our tight binding calculations show two distinct topological phases with the critical point at $\gamma_2=(t_1+t_2)$. For $\gamma_2<(t_1+t_2)$, the Zak phase is quantized as $3\pi$ and there are couple of  zero energy modes.
In the present study we did not focus on the intermediate stacking arrangements, which may be arrived by starting from either AA or AB and sliding another chain by a distance which in not an integer multiple of the half-unit cell length. The electronic structure of the model in those asymmetric stacking, as discussed in \cite{basu2022structural} suffers from the absence of particle-hole symmetry and does not fall in any of the topological classes for 1-d systems. 

As Polyyne chains have been experimentally synthesised \cite{cataldo2004synthesis,kutrovskaya2020excitonic}, these topological phases may be directly measured \cite{parto2018edge} %\textcolor{red}{look for experiments of detecting edge states in 1-d}.
Secondly, simulating the model by ultracold atoms \cite{atala2013direct}, Zak phases for the 3 different topological phases of the system can be experimentally demonstrated.

Apart from experimental realization, with the advent of quantum information (QI) theoretic techniques, probes like%to capture this  Another probe to study topological phases is the calculation of 
disconnected entanglement entropy \cite{zeng2019quantum,micallo2020topological} should be studied in a system like ours, because of the inherent richness of topological phases of it. This is tantalizing from QI perspective itself, in order to pin-point the exact entanglement measure which may classify quantum/topological dynamical phase transitions in periodic systems.
%which could distinguish topological and non-topological insulating phases in open boundary conditions. Here we do not study the system with the effect of any immediate sliding since charge conjugation symmetry is broken.

%and according to classification of topological insulators based on discrete symmetries such system is a trivial insulator.
\section*{Acknowledgements}
Research of RB is supported by the following grants from the SERB, India: CRG/2020/002035, SRG/2020/001037. Discussions with Indrakshi Raychowdhury and Raka Dasgupta are gratefully acknowledged.  
\appendix
\section*{Appendix}
\subsection{AB stacking eigenvectors}
The occupied eigen vectors for the AB stacking arrangement required for the calculation of the Zak phase are
\small{
\begin{equation*}
    \ket{{X}_{3}^{AB}}=
    \frac{1}{2}
    \begin{pmatrix}
     -\frac{e^{-iAk}\sqrt{2(1+\cos{Ak})({|f|}^2+{\gamma_2}^2-2{\gamma_2}(t_1+t_2)\cos{\frac{Ak}{2}}  ) } }{\gamma_2\sqrt{2(1+\cos{AK})}-((t_1+t_2)(1+\cos{Ak})-i(t_1-t_2)\sin{Ak})}\\
     -\frac{\sqrt{2(1+\cos{Ak})}}{1+e^{iAk}}\\
     \frac{(1+e^{-iAk})\sqrt{({|f|}^2+{\gamma_2}^2-2{\gamma_2}(t_1+t_2)\cos{\frac{Ak}{2}}  ) } }{\gamma_2\sqrt{2(1+\cos{AK})}-((t_1+t_2)(1+\cos{Ak})-i(t_1-t_2)\sin{Ak})} \\
     1\\
    \end{pmatrix},
\end{equation*}
\begin{equation}
    \ket{{X}_{4}^{AB}}=
    \frac{1}{2}
    \begin{pmatrix}
     -\frac{e^{-iAk}\sqrt{2(1+\cos{Ak})({|f|}^2+{\gamma_2}^2+{\gamma_2}(t_1+t_2)\sqrt{2(1+\cos{Ak})} }}{\gamma_2\sqrt{2(1+\cos{AK})}+(t_1+t_2)(1+\cos{Ak}-\frac{i(t_1-t_2)\sin{Ak}}{(t_1+t_2)}  )} \\
     \frac{\sqrt{2(1+\cos{Ak})}}{1+e^{iAk}}\\
    -\frac{(1+e^{-iAk})\sqrt{({|f|}^2+{\gamma_2}^2+{\gamma_2}(t_1+t_2)\sqrt{2(1+\cos{Ak})} ) } }{\gamma_2\sqrt{2(1+\cos{Ak})}+(t_1+t_2)(1+\cos{Ak}-\frac{i(t_1-t_2)\sin{Ak}}{(t_1+t_2)}  )} \\
     1\\
    \end{pmatrix}.
    \label{}
\end{equation}}
\subsection{AB stacking Hamiltonian kernel matrix}
We write the Hamiltonian kernel in \eqref{AB_h} in the real basis $\Psi^{\dag}=( b^{\dag}_1 \hspace{0.4cm} \tilde{a}^{\dag}_1 \hspace{0.4cm} \tilde{b}^{\dag}_p \hspace{0.4cm} a^{\dag}_{p+1} \hspace{0.3cm} \hdots \hspace{0.3cm} \tilde{b}^{\dag}_{P-1} \hspace{0.4cm} a^{\dag}_{P} )$
\begin{align}
\label{matrix}
H_{ab}=\begin{pmatrix}
 0 & \gamma_2 & 0 & t_2 & 0 & 0 & 0 & 0 & \hdots & 0 & 0 \\
 \gamma_2 & 0& t_1& 0 & 0 & 0 & 0 & 0 & \hdots & 0 & 0\\
 0 & t_1& 0& \gamma_2 & 0 & t_2 & 0 & 0 & \hdots & 0 & 0\\
 t_2 & 0& \gamma_2& 0 & t_1 & 0 & 0 & 0 & \hdots & 0 & 0\\
 0 & 0 & 0 & t_1 & 0 & \gamma_2 & 0 & t_2 & \hdots & 0 & 0\\
 0 & 0 & t_2 & 0 & \gamma_2 & 0 & t_1 & 0 & \hdots & 0 & 0\\
 \vdots&\vdots&\vdots&\vdots&\vdots&\vdots&\vdots&\vdots&\ddots&\vdots&\vdots&\\
 0 & 0 & 0 & 0 & 0 & 0 & 0 & 0 &\hdots & 0 & \gamma_2\\
 0 & 0 & 0 & 0 & 0 & 0 & 0 & 0 & \hdots & \gamma_2 & 0
\end{pmatrix}
.
\end{align}
\subsection{Convergence of the norm of the zero eigenvector}
\label{series}
From \eqref{ab_1}, we have the square of the norm of the zero mode eigenvector to be  $\sum_{n=1}^{N} (|B_n|^2+|D_n|^2)$, where:
\begin{equation}
     |B_n|^2= \left( \frac{{\gamma_2}^2-2t_1t_2}{2{t_1}^2}\right)^n 
     \left( C_1^2+C_2^2+2C_1C_2\cos \left( 2 \beta n \right)\right),
     \label{mod}
\end{equation}
\begin{equation}
    |D_n|^2= \left( \frac{{\gamma_2}^2-2t_1t_2}{2{t_2}^2}\right)^n 
    \left( C_3^2+C_4^2+2C_3C_4  \cos \left( 2 \beta n \right)\right).
\label{mod1}
\end{equation}
where $\beta=\tan ^{-1} (\frac{\sqrt{|{\gamma_2}^2-4t_1t_2}|}{\gamma_2})$.
Clearly, the necessary condition for the above sum to converge as $N \rightarrow \infty$ is  $(\frac{{\gamma_2}^2-2t_1t_2}{2{t_1}^2})<1$ and  $(\frac{{\gamma_2}^2-2t_1t_2}{2{t_2}^2})<1$. Both of these conditions are satisfied only if $\gamma_2 < t_1+t_2$. But, the equations \eqref{mod} and \eqref{mod1} are valid for $\gamma_2^2 < 4t_1 t_2 $ and $\gamma_2 < t_1+t_2$ already satisfies this. On the other hand, if we consider ${\gamma_2}^2>4t_1t_2$, the we have
\begin{equation}
|B_n|^2=\left( C_1\left(\frac{\gamma_2+\kappa}{2t_1}\right)^n +C_2\left(\frac{\gamma_2-\kappa}{2t_1}\right)^n \right)^2,
\label{gr}
\end{equation}
\begin{equation}
|D_n|^2=\left( C_3\left(\frac{\gamma_2+\kappa}{2t_2}\right)^n +C_4\left(\frac{\gamma_2-\kappa}{2t_2}\right)^n \right)^2.
\label{gr1}
\end{equation}
The necessary condition for the sum $\sum_{n=1}^{N} (|B_n|^2+|D_n|^2)$ to converge as $N\rightarrow\infty$  is $\gamma_2 <(t_1+t_2)$. 

\bibliography{main}

\providecommand{\noopsort}[1]{}\providecommand{\singleletter}[1]{#1}%
\begin{thebibliography}{38}%
\makeatletter
\providecommand \@ifxundefined [1]{%
 \@ifx{#1\undefined}
}%
\providecommand \@ifnum [1]{%
 \ifnum #1\expandafter \@firstoftwo
 \else \expandafter \@secondoftwo
 \fi
}%
\providecommand \@ifx [1]{%
 \ifx #1\expandafter \@firstoftwo
 \else \expandafter \@secondoftwo
 \fi
}%
\providecommand \natexlab [1]{#1}%
\providecommand \enquote  [1]{``#1''}%
\providecommand \bibnamefont  [1]{#1}%
\providecommand \bibfnamefont [1]{#1}%
\providecommand \citenamefont [1]{#1}%
\providecommand \href@noop [0]{\@secondoftwo}%
\providecommand \href [0]{\begingroup \@sanitize@url \@href}%
\providecommand \@href[1]{\@@startlink{#1}\@@href}%
\providecommand \@@href[1]{\endgroup#1\@@endlink}%
\providecommand \@sanitize@url [0]{\catcode `\\12\catcode `\$12\catcode
  `\&12\catcode `\#12\catcode `\^12\catcode `\_12\catcode `\%12\relax}%
\providecommand \@@startlink[1]{}%
\providecommand \@@endlink[0]{}%
\providecommand \url  [0]{\begingroup\@sanitize@url \@url }%
\providecommand \@url [1]{\endgroup\@href {#1}{\urlprefix }}%
\providecommand \urlprefix  [0]{URL }%
\providecommand \Eprint [0]{\href }%
\providecommand \doibase [0]{https://doi.org/}%
\providecommand \selectlanguage [0]{\@gobble}%
\providecommand \bibinfo  [0]{\@secondoftwo}%
\providecommand \bibfield  [0]{\@secondoftwo}%
\providecommand \translation [1]{[#1]}%
\providecommand \BibitemOpen [0]{}%
\providecommand \bibitemStop [0]{}%
\providecommand \bibitemNoStop [0]{.\EOS\space}%
\providecommand \EOS [0]{\spacefactor3000\relax}%
\providecommand \BibitemShut  [1]{\csname bibitem#1\endcsname}%
\let\auto@bib@innerbib\@empty
%</preamble>
\bibitem [{\citenamefont {Su}\ \emph {et~al.}(1979)\citenamefont {Su},
  \citenamefont {Schrieffer},\ and\ \citenamefont {Heeger}}]{su1979solitons}%
  \BibitemOpen
  \bibfield  {author} {\bibinfo {author} {\bibfnamefont {W.}~\bibnamefont
  {Su}}, \bibinfo {author} {\bibfnamefont {J.}~\bibnamefont {Schrieffer}},\
  and\ \bibinfo {author} {\bibfnamefont {A.~J.}\ \bibnamefont {Heeger}},\
  }\href@noop {} {\bibfield  {journal} {\bibinfo  {journal} {Physical review
  letters}\ }\textbf {\bibinfo {volume} {42}},\ \bibinfo {pages} {1698}
  (\bibinfo {year} {1979})}\BibitemShut {NoStop}%
\bibitem [{\citenamefont {Su}\ \emph {et~al.}(1980)\citenamefont {Su},
  \citenamefont {Schrieffer},\ and\ \citenamefont {Heeger}}]{su1980soliton}%
  \BibitemOpen
  \bibfield  {author} {\bibinfo {author} {\bibfnamefont {W.-P.}\ \bibnamefont
  {Su}}, \bibinfo {author} {\bibfnamefont {J.}~\bibnamefont {Schrieffer}},\
  and\ \bibinfo {author} {\bibfnamefont {A.}~\bibnamefont {Heeger}},\
  }\href@noop {} {\bibfield  {journal} {\bibinfo  {journal} {Physical Review
  B}\ }\textbf {\bibinfo {volume} {22}},\ \bibinfo {pages} {2099} (\bibinfo
  {year} {1980})}\BibitemShut {NoStop}%
\bibitem [{\citenamefont {Heeger}\ \emph {et~al.}(1988)\citenamefont {Heeger},
  \citenamefont {Kivelson}, \citenamefont {Schrieffer},\ and\ \citenamefont
  {Su}}]{heeger1988solitons}%
  \BibitemOpen
  \bibfield  {author} {\bibinfo {author} {\bibfnamefont {A.~J.}\ \bibnamefont
  {Heeger}}, \bibinfo {author} {\bibfnamefont {S.}~\bibnamefont {Kivelson}},
  \bibinfo {author} {\bibfnamefont {J.}~\bibnamefont {Schrieffer}},\ and\
  \bibinfo {author} {\bibfnamefont {W.-P.}\ \bibnamefont {Su}},\ }\href@noop {}
  {\bibfield  {journal} {\bibinfo  {journal} {Reviews of Modern Physics}\
  }\textbf {\bibinfo {volume} {60}},\ \bibinfo {pages} {781} (\bibinfo {year}
  {1988})}\BibitemShut {NoStop}%
\bibitem [{\citenamefont {Kitagawa}\ \emph {et~al.}(2012)\citenamefont
  {Kitagawa}, \citenamefont {Broome}, \citenamefont {Fedrizzi}, \citenamefont
  {Rudner}, \citenamefont {Berg}, \citenamefont {Kassal}, \citenamefont
  {Aspuru-Guzik}, \citenamefont {Demler},\ and\ \citenamefont
  {White}}]{kitagawa2012observation}%
  \BibitemOpen
  \bibfield  {author} {\bibinfo {author} {\bibfnamefont {T.}~\bibnamefont
  {Kitagawa}}, \bibinfo {author} {\bibfnamefont {M.~A.}\ \bibnamefont
  {Broome}}, \bibinfo {author} {\bibfnamefont {A.}~\bibnamefont {Fedrizzi}},
  \bibinfo {author} {\bibfnamefont {M.~S.}\ \bibnamefont {Rudner}}, \bibinfo
  {author} {\bibfnamefont {E.}~\bibnamefont {Berg}}, \bibinfo {author}
  {\bibfnamefont {I.}~\bibnamefont {Kassal}}, \bibinfo {author} {\bibfnamefont
  {A.}~\bibnamefont {Aspuru-Guzik}}, \bibinfo {author} {\bibfnamefont
  {E.}~\bibnamefont {Demler}},\ and\ \bibinfo {author} {\bibfnamefont {A.~G.}\
  \bibnamefont {White}},\ }\href@noop {} {\bibfield  {journal} {\bibinfo
  {journal} {Nature communications}\ }\textbf {\bibinfo {volume} {3}},\
  \bibinfo {pages} {1} (\bibinfo {year} {2012})}\BibitemShut {NoStop}%
\bibitem [{\citenamefont {Atala}\ \emph {et~al.}(2013)\citenamefont {Atala},
  \citenamefont {Aidelsburger}, \citenamefont {Barreiro}, \citenamefont
  {Abanin}, \citenamefont {Kitagawa}, \citenamefont {Demler},\ and\
  \citenamefont {Bloch}}]{atala2013direct}%
  \BibitemOpen
  \bibfield  {author} {\bibinfo {author} {\bibfnamefont {M.}~\bibnamefont
  {Atala}}, \bibinfo {author} {\bibfnamefont {M.}~\bibnamefont {Aidelsburger}},
  \bibinfo {author} {\bibfnamefont {J.~T.}\ \bibnamefont {Barreiro}}, \bibinfo
  {author} {\bibfnamefont {D.}~\bibnamefont {Abanin}}, \bibinfo {author}
  {\bibfnamefont {T.}~\bibnamefont {Kitagawa}}, \bibinfo {author}
  {\bibfnamefont {E.}~\bibnamefont {Demler}},\ and\ \bibinfo {author}
  {\bibfnamefont {I.}~\bibnamefont {Bloch}},\ }\href@noop {} {\bibfield
  {journal} {\bibinfo  {journal} {Nature Physics}\ }\textbf {\bibinfo {volume}
  {9}},\ \bibinfo {pages} {795} (\bibinfo {year} {2013})}\BibitemShut {NoStop}%
\bibitem [{\citenamefont {Nakajima}\ \emph {et~al.}(2016)\citenamefont
  {Nakajima}, \citenamefont {Tomita}, \citenamefont {Taie}, \citenamefont
  {Ichinose}, \citenamefont {Ozawa}, \citenamefont {Wang}, \citenamefont
  {Troyer},\ and\ \citenamefont {Takahashi}}]{nakajima2016topological}%
  \BibitemOpen
  \bibfield  {author} {\bibinfo {author} {\bibfnamefont {S.}~\bibnamefont
  {Nakajima}}, \bibinfo {author} {\bibfnamefont {T.}~\bibnamefont {Tomita}},
  \bibinfo {author} {\bibfnamefont {S.}~\bibnamefont {Taie}}, \bibinfo {author}
  {\bibfnamefont {T.}~\bibnamefont {Ichinose}}, \bibinfo {author}
  {\bibfnamefont {H.}~\bibnamefont {Ozawa}}, \bibinfo {author} {\bibfnamefont
  {L.}~\bibnamefont {Wang}}, \bibinfo {author} {\bibfnamefont {M.}~\bibnamefont
  {Troyer}},\ and\ \bibinfo {author} {\bibfnamefont {Y.}~\bibnamefont
  {Takahashi}},\ }\href@noop {} {\bibfield  {journal} {\bibinfo  {journal}
  {Nature Physics}\ }\textbf {\bibinfo {volume} {12}},\ \bibinfo {pages} {296}
  (\bibinfo {year} {2016})}\BibitemShut {NoStop}%
\bibitem [{\citenamefont {Lohse}\ \emph {et~al.}(2016)\citenamefont {Lohse},
  \citenamefont {Schweizer}, \citenamefont {Zilberberg}, \citenamefont
  {Aidelsburger},\ and\ \citenamefont {Bloch}}]{lohse2016thouless}%
  \BibitemOpen
  \bibfield  {author} {\bibinfo {author} {\bibfnamefont {M.}~\bibnamefont
  {Lohse}}, \bibinfo {author} {\bibfnamefont {C.}~\bibnamefont {Schweizer}},
  \bibinfo {author} {\bibfnamefont {O.}~\bibnamefont {Zilberberg}}, \bibinfo
  {author} {\bibfnamefont {M.}~\bibnamefont {Aidelsburger}},\ and\ \bibinfo
  {author} {\bibfnamefont {I.}~\bibnamefont {Bloch}},\ }\href@noop {}
  {\bibfield  {journal} {\bibinfo  {journal} {Nature Physics}\ }\textbf
  {\bibinfo {volume} {12}},\ \bibinfo {pages} {350} (\bibinfo {year}
  {2016})}\BibitemShut {NoStop}%
\bibitem [{\citenamefont {Xie}\ \emph {et~al.}(2018)\citenamefont {Xie},
  \citenamefont {Wang}, \citenamefont {Wang}, \citenamefont {Zhu},
  \citenamefont {Jiang}, \citenamefont {Lu},\ and\ \citenamefont
  {Chen}}]{xie2018second}%
  \BibitemOpen
  \bibfield  {author} {\bibinfo {author} {\bibfnamefont {B.-Y.}\ \bibnamefont
  {Xie}}, \bibinfo {author} {\bibfnamefont {H.-F.}\ \bibnamefont {Wang}},
  \bibinfo {author} {\bibfnamefont {H.-X.}\ \bibnamefont {Wang}}, \bibinfo
  {author} {\bibfnamefont {X.-Y.}\ \bibnamefont {Zhu}}, \bibinfo {author}
  {\bibfnamefont {J.-H.}\ \bibnamefont {Jiang}}, \bibinfo {author}
  {\bibfnamefont {M.-H.}\ \bibnamefont {Lu}},\ and\ \bibinfo {author}
  {\bibfnamefont {Y.-F.}\ \bibnamefont {Chen}},\ }\href@noop {} {\bibfield
  {journal} {\bibinfo  {journal} {Physical Review B}\ }\textbf {\bibinfo
  {volume} {98}},\ \bibinfo {pages} {205147} (\bibinfo {year}
  {2018})}\BibitemShut {NoStop}%
\bibitem [{\citenamefont {Asb{\'o}th}\ \emph {et~al.}(2016)\citenamefont
  {Asb{\'o}th}, \citenamefont {Oroszl{\'a}ny},\ and\ \citenamefont
  {P{\'a}lyi}}]{asboth2016short}%
  \BibitemOpen
  \bibfield  {author} {\bibinfo {author} {\bibfnamefont {J.~K.}\ \bibnamefont
  {Asb{\'o}th}}, \bibinfo {author} {\bibfnamefont {L.}~\bibnamefont
  {Oroszl{\'a}ny}},\ and\ \bibinfo {author} {\bibfnamefont {A.}~\bibnamefont
  {P{\'a}lyi}},\ }\href@noop {} {\bibfield  {journal} {\bibinfo  {journal}
  {Lecture notes in physics}\ }\textbf {\bibinfo {volume} {919}},\ \bibinfo
  {pages} {166} (\bibinfo {year} {2016})}\BibitemShut {NoStop}%
\bibitem [{\citenamefont {Li}\ \emph {et~al.}(2014)\citenamefont {Li},
  \citenamefont {Xu},\ and\ \citenamefont {Chen}}]{li2014topological}%
  \BibitemOpen
  \bibfield  {author} {\bibinfo {author} {\bibfnamefont {L.}~\bibnamefont
  {Li}}, \bibinfo {author} {\bibfnamefont {Z.}~\bibnamefont {Xu}},\ and\
  \bibinfo {author} {\bibfnamefont {S.}~\bibnamefont {Chen}},\ }\href@noop {}
  {\bibfield  {journal} {\bibinfo  {journal} {Physical Review B}\ }\textbf
  {\bibinfo {volume} {89}},\ \bibinfo {pages} {085111} (\bibinfo {year}
  {2014})}\BibitemShut {NoStop}%
\bibitem [{\citenamefont {Delplace}\ \emph {et~al.}(2011)\citenamefont
  {Delplace}, \citenamefont {Ullmo},\ and\ \citenamefont
  {Montambaux}}]{delplace2011zak}%
  \BibitemOpen
  \bibfield  {author} {\bibinfo {author} {\bibfnamefont {P.}~\bibnamefont
  {Delplace}}, \bibinfo {author} {\bibfnamefont {D.}~\bibnamefont {Ullmo}},\
  and\ \bibinfo {author} {\bibfnamefont {G.}~\bibnamefont {Montambaux}},\
  }\href@noop {} {\bibfield  {journal} {\bibinfo  {journal} {Physical Review
  B}\ }\textbf {\bibinfo {volume} {84}},\ \bibinfo {pages} {195452} (\bibinfo
  {year} {2011})}\BibitemShut {NoStop}%
\bibitem [{\citenamefont {Hasan}\ and\ \citenamefont
  {Kane}(2010)}]{hasan2010colloquium}%
  \BibitemOpen
  \bibfield  {author} {\bibinfo {author} {\bibfnamefont {M.~Z.}\ \bibnamefont
  {Hasan}}\ and\ \bibinfo {author} {\bibfnamefont {C.~L.}\ \bibnamefont
  {Kane}},\ }\href@noop {} {\bibfield  {journal} {\bibinfo  {journal} {Reviews
  of modern physics}\ }\textbf {\bibinfo {volume} {82}},\ \bibinfo {pages}
  {3045} (\bibinfo {year} {2010})}\BibitemShut {NoStop}%
\bibitem [{\citenamefont {Qi}\ and\ \citenamefont
  {Zhang}(2011)}]{qi2011topological}%
  \BibitemOpen
  \bibfield  {author} {\bibinfo {author} {\bibfnamefont {X.-L.}\ \bibnamefont
  {Qi}}\ and\ \bibinfo {author} {\bibfnamefont {S.-C.}\ \bibnamefont {Zhang}},\
  }\href@noop {} {\bibfield  {journal} {\bibinfo  {journal} {Reviews of Modern
  Physics}\ }\textbf {\bibinfo {volume} {83}},\ \bibinfo {pages} {1057}
  (\bibinfo {year} {2011})}\BibitemShut {NoStop}%
\bibitem [{\citenamefont {Basu}\ and\ \citenamefont
  {Bhattacharyya}(2022)}]{basu2022structural}%
  \BibitemOpen
  \bibfield  {author} {\bibinfo {author} {\bibfnamefont {R.}~\bibnamefont
  {Basu}}\ and\ \bibinfo {author} {\bibfnamefont {S.}~\bibnamefont
  {Bhattacharyya}},\ }\href@noop {} {\bibfield  {journal} {\bibinfo  {journal}
  {Carbon Trends}\ }\textbf {\bibinfo {volume} {7}},\ \bibinfo {pages} {100163}
  (\bibinfo {year} {2022})}\BibitemShut {NoStop}%
\bibitem [{\citenamefont {Charlier}\ \emph {et~al.}(1992)\citenamefont
  {Charlier}, \citenamefont {Michenaud},\ and\ \citenamefont
  {Gonze}}]{charlier1992first}%
  \BibitemOpen
  \bibfield  {author} {\bibinfo {author} {\bibfnamefont {J.-C.}\ \bibnamefont
  {Charlier}}, \bibinfo {author} {\bibfnamefont {J.-P.}\ \bibnamefont
  {Michenaud}},\ and\ \bibinfo {author} {\bibfnamefont {X.}~\bibnamefont
  {Gonze}},\ }\href@noop {} {\bibfield  {journal} {\bibinfo  {journal}
  {Physical Review B}\ }\textbf {\bibinfo {volume} {46}},\ \bibinfo {pages}
  {4531} (\bibinfo {year} {1992})}\BibitemShut {NoStop}%
\bibitem [{\citenamefont {Kitaev}(2009)}]{kitaev2009periodic}%
  \BibitemOpen
  \bibfield  {author} {\bibinfo {author} {\bibfnamefont {A.}~\bibnamefont
  {Kitaev}}\ }(\bibinfo {organization} {American Institute of Physics},\
  \bibinfo {year} {2009})\ pp.\ \bibinfo {pages} {22--30}\BibitemShut {NoStop}%
\bibitem [{\citenamefont {Chiu}\ \emph {et~al.}(2013)\citenamefont {Chiu},
  \citenamefont {Yao},\ and\ \citenamefont {Ryu}}]{chiu2013classification}%
  \BibitemOpen
  \bibfield  {author} {\bibinfo {author} {\bibfnamefont {C.-K.}\ \bibnamefont
  {Chiu}}, \bibinfo {author} {\bibfnamefont {H.}~\bibnamefont {Yao}},\ and\
  \bibinfo {author} {\bibfnamefont {S.}~\bibnamefont {Ryu}},\ }\href@noop {}
  {\bibfield  {journal} {\bibinfo  {journal} {Physical Review B}\ }\textbf
  {\bibinfo {volume} {88}},\ \bibinfo {pages} {075142} (\bibinfo {year}
  {2013})}\BibitemShut {NoStop}%
\bibitem [{\citenamefont {Zak}(1989)}]{zak1989berry}%
  \BibitemOpen
  \bibfield  {author} {\bibinfo {author} {\bibfnamefont {J.}~\bibnamefont
  {Zak}},\ }\href@noop {} {\bibfield  {journal} {\bibinfo  {journal} {Physical
  review letters}\ }\textbf {\bibinfo {volume} {62}},\ \bibinfo {pages} {2747}
  (\bibinfo {year} {1989})}\BibitemShut {NoStop}%
\bibitem [{\citenamefont {Ryu}\ and\ \citenamefont
  {Hatsugai}(2002)}]{ryu2002topological}%
  \BibitemOpen
  \bibfield  {author} {\bibinfo {author} {\bibfnamefont {S.}~\bibnamefont
  {Ryu}}\ and\ \bibinfo {author} {\bibfnamefont {Y.}~\bibnamefont {Hatsugai}},\
  }\href@noop {} {\bibfield  {journal} {\bibinfo  {journal} {Physical review
  letters}\ }\textbf {\bibinfo {volume} {89}},\ \bibinfo {pages} {077002}
  (\bibinfo {year} {2002})}\BibitemShut {NoStop}%
\bibitem [{\citenamefont {Lang}\ \emph {et~al.}(2012)\citenamefont {Lang},
  \citenamefont {Cai},\ and\ \citenamefont {Chen}}]{lang2012edge}%
  \BibitemOpen
  \bibfield  {author} {\bibinfo {author} {\bibfnamefont {L.-J.}\ \bibnamefont
  {Lang}}, \bibinfo {author} {\bibfnamefont {X.}~\bibnamefont {Cai}},\ and\
  \bibinfo {author} {\bibfnamefont {S.}~\bibnamefont {Chen}},\ }\href@noop {}
  {\bibfield  {journal} {\bibinfo  {journal} {Physical review letters}\
  }\textbf {\bibinfo {volume} {108}},\ \bibinfo {pages} {220401} (\bibinfo
  {year} {2012})}\BibitemShut {NoStop}%
\bibitem [{\citenamefont {Rusznyak}\ \emph {et~al.}(2005)\citenamefont
  {Rusznyak}, \citenamefont {Z{\'o}lyomi}, \citenamefont {K{\"u}rti},
  \citenamefont {Yang},\ and\ \citenamefont {Kertesz}}]{rusznyak2005bond}%
  \BibitemOpen
  \bibfield  {author} {\bibinfo {author} {\bibfnamefont {A.}~\bibnamefont
  {Rusznyak}}, \bibinfo {author} {\bibfnamefont {V.}~\bibnamefont
  {Z{\'o}lyomi}}, \bibinfo {author} {\bibfnamefont {J.}~\bibnamefont
  {K{\"u}rti}}, \bibinfo {author} {\bibfnamefont {S.}~\bibnamefont {Yang}},\
  and\ \bibinfo {author} {\bibfnamefont {M.}~\bibnamefont {Kertesz}},\
  }\href@noop {} {\bibfield  {journal} {\bibinfo  {journal} {Physical Review
  B}\ }\textbf {\bibinfo {volume} {72}},\ \bibinfo {pages} {155420} (\bibinfo
  {year} {2005})}\BibitemShut {NoStop}%
\bibitem [{\citenamefont {Lambropoulos}\ and\ \citenamefont
  {Simserides}(2017)}]{lambropoulos2017electronic}%
  \BibitemOpen
  \bibfield  {author} {\bibinfo {author} {\bibfnamefont {K.}~\bibnamefont
  {Lambropoulos}}\ and\ \bibinfo {author} {\bibfnamefont {C.}~\bibnamefont
  {Simserides}},\ }\href@noop {} {\bibfield  {journal} {\bibinfo  {journal}
  {Physical Chemistry Chemical Physics}\ }\textbf {\bibinfo {volume} {19}},\
  \bibinfo {pages} {26890} (\bibinfo {year} {2017})}\BibitemShut {NoStop}%
\bibitem [{\citenamefont {Al-Backri}\ \emph {et~al.}(2014)\citenamefont
  {Al-Backri}, \citenamefont {Z{\'o}lyomi},\ and\ \citenamefont
  {Lambert}}]{al2014electronic}%
  \BibitemOpen
  \bibfield  {author} {\bibinfo {author} {\bibfnamefont {A.}~\bibnamefont
  {Al-Backri}}, \bibinfo {author} {\bibfnamefont {V.}~\bibnamefont
  {Z{\'o}lyomi}},\ and\ \bibinfo {author} {\bibfnamefont {C.~J.}\ \bibnamefont
  {Lambert}},\ }\href@noop {} {\bibfield  {journal} {\bibinfo  {journal} {The
  Journal of chemical physics}\ }\textbf {\bibinfo {volume} {140}},\ \bibinfo
  {pages} {104306} (\bibinfo {year} {2014})}\BibitemShut {NoStop}%
\bibitem [{\citenamefont {Kartoon}\ \emph {et~al.}(2018)\citenamefont
  {Kartoon}, \citenamefont {Argaman},\ and\ \citenamefont
  {Makov}}]{kartoon2018driving}%
  \BibitemOpen
  \bibfield  {author} {\bibinfo {author} {\bibfnamefont {D.}~\bibnamefont
  {Kartoon}}, \bibinfo {author} {\bibfnamefont {U.}~\bibnamefont {Argaman}},\
  and\ \bibinfo {author} {\bibfnamefont {G.}~\bibnamefont {Makov}},\
  }\href@noop {} {\bibfield  {journal} {\bibinfo  {journal} {Physical Review
  B}\ }\textbf {\bibinfo {volume} {98}},\ \bibinfo {pages} {165429} (\bibinfo
  {year} {2018})}\BibitemShut {NoStop}%
\bibitem [{\citenamefont {Padavi{\'c}}\ \emph {et~al.}(2018)\citenamefont
  {Padavi{\'c}}, \citenamefont {Hegde}, \citenamefont {DeGottardi},\ and\
  \citenamefont {Vishveshwara}}]{padavic2018topological}%
  \BibitemOpen
  \bibfield  {author} {\bibinfo {author} {\bibfnamefont {K.}~\bibnamefont
  {Padavi{\'c}}}, \bibinfo {author} {\bibfnamefont {S.~S.}\ \bibnamefont
  {Hegde}}, \bibinfo {author} {\bibfnamefont {W.}~\bibnamefont {DeGottardi}},\
  and\ \bibinfo {author} {\bibfnamefont {S.}~\bibnamefont {Vishveshwara}},\
  }\href@noop {} {\bibfield  {journal} {\bibinfo  {journal} {Physical Review
  B}\ }\textbf {\bibinfo {volume} {98}},\ \bibinfo {pages} {024205} (\bibinfo
  {year} {2018})}\BibitemShut {NoStop}%
\bibitem [{\citenamefont {Kurzyna}\ and\ \citenamefont
  {Kwapi{\'n}ski}(2020)}]{kurzyna2020edge}%
  \BibitemOpen
  \bibfield  {author} {\bibinfo {author} {\bibfnamefont {M.}~\bibnamefont
  {Kurzyna}}\ and\ \bibinfo {author} {\bibfnamefont {T.}~\bibnamefont
  {Kwapi{\'n}ski}},\ }\href@noop {} {\bibfield  {journal} {\bibinfo  {journal}
  {Physical Review B}\ }\textbf {\bibinfo {volume} {102}},\ \bibinfo {pages}
  {195429} (\bibinfo {year} {2020})}\BibitemShut {NoStop}%
\bibitem [{\citenamefont {Jangjan}\ and\ \citenamefont
  {Hosseini}(2020)}]{jangjan2020floquet}%
  \BibitemOpen
  \bibfield  {author} {\bibinfo {author} {\bibfnamefont {M.}~\bibnamefont
  {Jangjan}}\ and\ \bibinfo {author} {\bibfnamefont {M.~V.}\ \bibnamefont
  {Hosseini}},\ }\href@noop {} {\bibfield  {journal} {\bibinfo  {journal}
  {Scientific Reports}\ }\textbf {\bibinfo {volume} {10}},\ \bibinfo {pages}
  {1} (\bibinfo {year} {2020})}\BibitemShut {NoStop}%
\bibitem [{\citenamefont {Kariyado}\ and\ \citenamefont
  {Vishwanath}(2019)}]{kariyado2019flat}%
  \BibitemOpen
  \bibfield  {author} {\bibinfo {author} {\bibfnamefont {T.}~\bibnamefont
  {Kariyado}}\ and\ \bibinfo {author} {\bibfnamefont {A.}~\bibnamefont
  {Vishwanath}},\ }\href@noop {} {\bibfield  {journal} {\bibinfo  {journal}
  {Physical Review Research}\ }\textbf {\bibinfo {volume} {1}},\ \bibinfo
  {pages} {033076} (\bibinfo {year} {2019})}\BibitemShut {NoStop}%
\bibitem [{\citenamefont {Luo}\ \emph {et~al.}(2022)\citenamefont {Luo},
  \citenamefont {Guan}, \citenamefont {Fan}, \citenamefont {Chen},\ and\
  \citenamefont {Jia}}]{luo2022topological}%
  \BibitemOpen
  \bibfield  {author} {\bibinfo {author} {\bibfnamefont {T.}~\bibnamefont
  {Luo}}, \bibinfo {author} {\bibfnamefont {X.}~\bibnamefont {Guan}}, \bibinfo
  {author} {\bibfnamefont {J.}~\bibnamefont {Fan}}, \bibinfo {author}
  {\bibfnamefont {G.}~\bibnamefont {Chen}},\ and\ \bibinfo {author}
  {\bibfnamefont {S.-T.}\ \bibnamefont {Jia}},\ }\href@noop {} {\bibfield
  {journal} {\bibinfo  {journal} {Chinese Physics B}\ }\textbf {\bibinfo
  {volume} {31}},\ \bibinfo {pages} {014208} (\bibinfo {year}
  {2022})}\BibitemShut {NoStop}%
\bibitem [{\citenamefont {Li}\ \emph {et~al.}(2017)\citenamefont {Li},
  \citenamefont {Lin}, \citenamefont {Zhang},\ and\ \citenamefont
  {Song}}]{li2017topological}%
  \BibitemOpen
  \bibfield  {author} {\bibinfo {author} {\bibfnamefont {C.}~\bibnamefont
  {Li}}, \bibinfo {author} {\bibfnamefont {S.}~\bibnamefont {Lin}}, \bibinfo
  {author} {\bibfnamefont {G.}~\bibnamefont {Zhang}},\ and\ \bibinfo {author}
  {\bibfnamefont {Z.}~\bibnamefont {Song}},\ }\href@noop {} {\bibfield
  {journal} {\bibinfo  {journal} {Physical Review B}\ }\textbf {\bibinfo
  {volume} {96}},\ \bibinfo {pages} {125418} (\bibinfo {year}
  {2017})}\BibitemShut {NoStop}%
\bibitem [{\citenamefont {Giannozzi}\ \emph {et~al.}(2009)\citenamefont
  {Giannozzi}, \citenamefont {Baroni}, \citenamefont {Bonini}, \citenamefont
  {Calandra}, \citenamefont {Car}, \citenamefont {Cavazzoni}, \citenamefont
  {Ceresoli}, \citenamefont {Chiarotti}, \citenamefont {Cococcioni},
  \citenamefont {Dabo} \emph {et~al.}}]{giannozzi2009quantum}%
  \BibitemOpen
  \bibfield  {author} {\bibinfo {author} {\bibfnamefont {P.}~\bibnamefont
  {Giannozzi}}, \bibinfo {author} {\bibfnamefont {S.}~\bibnamefont {Baroni}},
  \bibinfo {author} {\bibfnamefont {N.}~\bibnamefont {Bonini}}, \bibinfo
  {author} {\bibfnamefont {M.}~\bibnamefont {Calandra}}, \bibinfo {author}
  {\bibfnamefont {R.}~\bibnamefont {Car}}, \bibinfo {author} {\bibfnamefont
  {C.}~\bibnamefont {Cavazzoni}}, \bibinfo {author} {\bibfnamefont
  {D.}~\bibnamefont {Ceresoli}}, \bibinfo {author} {\bibfnamefont {G.~L.}\
  \bibnamefont {Chiarotti}}, \bibinfo {author} {\bibfnamefont {M.}~\bibnamefont
  {Cococcioni}}, \bibinfo {author} {\bibfnamefont {I.}~\bibnamefont {Dabo}},
  \emph {et~al.},\ }\href@noop {} {\bibfield  {journal} {\bibinfo  {journal}
  {Journal of physics: Condensed matter}\ }\textbf {\bibinfo {volume} {21}},\
  \bibinfo {pages} {395502} (\bibinfo {year} {2009})}\BibitemShut {NoStop}%
\bibitem [{\citenamefont {Midya}\ and\ \citenamefont
  {Feng}(2018)}]{midya2018topological}%
  \BibitemOpen
  \bibfield  {author} {\bibinfo {author} {\bibfnamefont {B.}~\bibnamefont
  {Midya}}\ and\ \bibinfo {author} {\bibfnamefont {L.}~\bibnamefont {Feng}},\
  }\href@noop {} {\bibfield  {journal} {\bibinfo  {journal} {Physical Review
  A}\ }\textbf {\bibinfo {volume} {98}},\ \bibinfo {pages} {043838} (\bibinfo
  {year} {2018})}\BibitemShut {NoStop}%
\bibitem [{\citenamefont {Lee}\ \emph {et~al.}(2022)\citenamefont {Lee},
  \citenamefont {Io},\ and\ \citenamefont {Kao}}]{lee2022winding}%
  \BibitemOpen
  \bibfield  {author} {\bibinfo {author} {\bibfnamefont {C.-S.}\ \bibnamefont
  {Lee}}, \bibinfo {author} {\bibfnamefont {I.-F.}\ \bibnamefont {Io}},\ and\
  \bibinfo {author} {\bibfnamefont {H.-c.}\ \bibnamefont {Kao}},\ }\href@noop
  {} {\bibfield  {journal} {\bibinfo  {journal} {Chinese Journal of Physics}\ }
  (\bibinfo {year} {2022})}\BibitemShut {NoStop}%
\bibitem [{\citenamefont {Cataldo}(2004)}]{cataldo2004synthesis}%
  \BibitemOpen
  \bibfield  {author} {\bibinfo {author} {\bibfnamefont {F.}~\bibnamefont
  {Cataldo}},\ }\href@noop {} {\bibfield  {journal} {\bibinfo  {journal}
  {Carbon}\ }\textbf {\bibinfo {volume} {42}},\ \bibinfo {pages} {129}
  (\bibinfo {year} {2004})}\BibitemShut {NoStop}%
\bibitem [{\citenamefont {Kutrovskaya}\ \emph {et~al.}(2020)\citenamefont
  {Kutrovskaya}, \citenamefont {Osipov}, \citenamefont {Baryshev},
  \citenamefont {Zasedatelev}, \citenamefont {Samyshkin}, \citenamefont
  {Demirchyan}, \citenamefont {Pulci}, \citenamefont {Grassano}, \citenamefont
  {Gontrani}, \citenamefont {Hartmann} \emph
  {et~al.}}]{kutrovskaya2020excitonic}%
  \BibitemOpen
  \bibfield  {author} {\bibinfo {author} {\bibfnamefont {S.}~\bibnamefont
  {Kutrovskaya}}, \bibinfo {author} {\bibfnamefont {A.}~\bibnamefont {Osipov}},
  \bibinfo {author} {\bibfnamefont {S.}~\bibnamefont {Baryshev}}, \bibinfo
  {author} {\bibfnamefont {A.}~\bibnamefont {Zasedatelev}}, \bibinfo {author}
  {\bibfnamefont {V.}~\bibnamefont {Samyshkin}}, \bibinfo {author}
  {\bibfnamefont {S.}~\bibnamefont {Demirchyan}}, \bibinfo {author}
  {\bibfnamefont {O.}~\bibnamefont {Pulci}}, \bibinfo {author} {\bibfnamefont
  {D.}~\bibnamefont {Grassano}}, \bibinfo {author} {\bibfnamefont
  {L.}~\bibnamefont {Gontrani}}, \bibinfo {author} {\bibfnamefont {R.~R.}\
  \bibnamefont {Hartmann}}, \emph {et~al.},\ }\href@noop {} {\bibfield
  {journal} {\bibinfo  {journal} {Nano letters}\ }\textbf {\bibinfo {volume}
  {20}},\ \bibinfo {pages} {6502} (\bibinfo {year} {2020})}\BibitemShut
  {NoStop}%
\bibitem [{\citenamefont {Parto}\ \emph {et~al.}(2018)\citenamefont {Parto},
  \citenamefont {Wittek}, \citenamefont {Hodaei}, \citenamefont {Harari},
  \citenamefont {Bandres}, \citenamefont {Ren}, \citenamefont {Rechtsman},
  \citenamefont {Segev}, \citenamefont {Christodoulides},\ and\ \citenamefont
  {Khajavikhan}}]{parto2018edge}%
  \BibitemOpen
  \bibfield  {author} {\bibinfo {author} {\bibfnamefont {M.}~\bibnamefont
  {Parto}}, \bibinfo {author} {\bibfnamefont {S.}~\bibnamefont {Wittek}},
  \bibinfo {author} {\bibfnamefont {H.}~\bibnamefont {Hodaei}}, \bibinfo
  {author} {\bibfnamefont {G.}~\bibnamefont {Harari}}, \bibinfo {author}
  {\bibfnamefont {M.~A.}\ \bibnamefont {Bandres}}, \bibinfo {author}
  {\bibfnamefont {J.}~\bibnamefont {Ren}}, \bibinfo {author} {\bibfnamefont
  {M.~C.}\ \bibnamefont {Rechtsman}}, \bibinfo {author} {\bibfnamefont
  {M.}~\bibnamefont {Segev}}, \bibinfo {author} {\bibfnamefont {D.~N.}\
  \bibnamefont {Christodoulides}},\ and\ \bibinfo {author} {\bibfnamefont
  {M.}~\bibnamefont {Khajavikhan}},\ }\href@noop {} {\bibfield  {journal}
  {\bibinfo  {journal} {Physical review letters}\ }\textbf {\bibinfo {volume}
  {120}},\ \bibinfo {pages} {113901} (\bibinfo {year} {2018})}\BibitemShut
  {NoStop}%
\bibitem [{\citenamefont {Zeng}\ \emph {et~al.}(2019)\citenamefont {Zeng},
  \citenamefont {Chen}, \citenamefont {Zhou}, \citenamefont {Wen} \emph
  {et~al.}}]{zeng2019quantum}%
  \BibitemOpen
  \bibfield  {author} {\bibinfo {author} {\bibfnamefont {B.}~\bibnamefont
  {Zeng}}, \bibinfo {author} {\bibfnamefont {X.}~\bibnamefont {Chen}}, \bibinfo
  {author} {\bibfnamefont {D.-L.}\ \bibnamefont {Zhou}}, \bibinfo {author}
  {\bibfnamefont {X.-G.}\ \bibnamefont {Wen}}, \emph {et~al.},\ }\href@noop {}
  {\emph {\bibinfo {title} {Quantum information meets quantum matter}}}\
  (\bibinfo  {publisher} {Springer},\ \bibinfo {year} {2019})\BibitemShut
  {NoStop}%
\bibitem [{\citenamefont {Micallo}\ \emph {et~al.}(2020)\citenamefont
  {Micallo}, \citenamefont {Vitale}, \citenamefont {Dalmonte},\ and\
  \citenamefont {Fromholz}}]{micallo2020topological}%
  \BibitemOpen
  \bibfield  {author} {\bibinfo {author} {\bibfnamefont {T.}~\bibnamefont
  {Micallo}}, \bibinfo {author} {\bibfnamefont {V.}~\bibnamefont {Vitale}},
  \bibinfo {author} {\bibfnamefont {M.}~\bibnamefont {Dalmonte}},\ and\
  \bibinfo {author} {\bibfnamefont {P.}~\bibnamefont {Fromholz}},\ }\href@noop
  {} {\bibfield  {journal} {\bibinfo  {journal} {SciPost Physics Core}\
  }\textbf {\bibinfo {volume} {3}},\ \bibinfo {pages} {012} (\bibinfo {year}
  {2020})}\BibitemShut {NoStop}%
\end{thebibliography}%
\end{document}